# Probing two-qubit capacitive interactions beyond bilinear regime using dual Hamiltonian parameter estimations

Jonginn Yun[1][†], Jaemin Park[1][†], Hyeongyu Jang[1], Jehyun Kim[1], Wonjin Jang[1], Youngwook Song[1], Min-Kyun Cho[1], Hanseo Sohn[1], Hwanchul Jung[2], Vladimir Umansky[3], and Dohun Kim[1]*

[1]Department of Physics and Astronomy, and Institute of Applied Physics, Seoul National University, Seoul 08826, Korea

[2] Department of Physics, Pusan National University, Busan 46241, Korea

[3]Braun Center for Submicron Research, Department of Condensed Matter Physics, Weizmann Institute of Science, Rehovot 76100, Israel

[†]These authors contributed equally to this work.

*Corresponding author: dohunkim@snu.ac.kr

**Abstract**

We report the simultaneous operation and two-qubit coupling measurement of a pair of two-electron spin qubits, actively decoupled from quasistatic nuclear noise in a GaAs quadruple quantum dot array. Coherent Rabi oscillations of both qubits (decay time ≈2 μs; frequency few MHz) are achieved by continuously tuning their drive frequency using rapidly converging real-time Hamiltonian estimators. We observe strong two-qubit capacitive interaction (> 190 MHz), combined with detuning pulses, inducing a state-conditional frequency shift. The two-qubit capacitive interaction is beyond the bilinear regime, consistent with recent theoretical predictions. We observe a high ratio (>16) between coherence and conditional phase-flip time, which supports the possibility of generating high-fidelity and fast quantum entanglement between encoded spin qubits using a simple capacitive interaction.

**Introduction**

Spins in semiconductor quantum dot (QD) nanostructures offer a promising platform for realizing scalable quantum information-processing units with high-fidelity universal quantum control[1–3]. Recent progress in III-V and IV semiconducting materials demonstrated several achievements including the demonstration of single- and two-qubit gate fidelities exceeding 99% in $^{Nat}$Si and $^{28}$Si (Ref. [2,3]), simultaneous qubit operations in GaAs with a coherence time over 2 µs (Ref. [4,5]), a few qubit entanglement in Ge and $^{28}$Si (Ref. [6–8]), high-temperature operations of spin qubits[9–11], and long-range coupling of spin qubits using superconducting cavity structures[12,13]. The field is currently moving towards the high-fidelity control of multiple qubits and the generation of controlled entanglement.

However, low-frequency noise, including quasi-static nuclear fluctuation and slow charge noise, is one of the main factors reducing coherence times below the intrinsic limit of a given host material[14–20]. For example, the spin coherence time is often affected by charge noise coupled through the inhomogeneous magnetic field generated by micromagnets[18,19]. Additionally, exchange- or capacitive-coupling-based two-qubit control is inherently susceptible to charge noise[14,16,20–22]. Thus, eliminating or mitigating slow magnetic and electric noises is an important task in semiconductor QD platforms.

Real-time Hamiltonian parameter estimation and measurement-based feedback[4,23,24] are two complementary techniques to coherent quantum feedback[25] capable of error mitigation, which is compatible - albeit sequentially - with general qubit controls. Previously, the Hamiltonian parameter estimation applied to GaAs has shown that the effect of quasi-static nuclear spin fluctuations can be strongly suppressed for single-spin[23] and singlet-triplet ($ST_0$) qubits[4]. Nevertheless, extending the technique to a multiqubit system is desirable.

While this has not been demonstrated to date, the simultaneous Hamiltonian estimation is also crucial for the accurate measurement of inter-qubit coupling strength. This is particularly important in the case of GaAs, for which the application of real-time calibrated single-qubit rotations on each qubit is a prerequisite.

In this study, we demonstrate the simultaneous drive of a pair of ST$_0$ qubits in GaAs. The quasi-static nuclear noise for each qubit is actively decoupled using a Bayesian inference-based real-time Hamiltonian estimation circuit. We show high-quality Rabi oscillations for both qubits with an oscillation quality factor above 10. We further exploit this result to demonstrate the measurement of the electrostatic coupling of two ST$_0$ qubits, which grows beyond an empirically assumed bilinear form[26,27] for large intra-qubit exchange energies. Combining this with the spin-echo sequence, we assess the potential to generate a high-fidelity and fast conditional phase gate using capacitive interactions in a linear QD array.

**Results**

**Simultaneous Hamiltonian parameter estimation**

Fig. 1a shows a quadruple QD device on top of a GaAs/AlGaAs heterostructure, hosting a pair of ST$_0$ qubits with singlet $|S\rangle$ and triplet-zero $|T_0\rangle$ basis states (See methods section and Ref.[28] for details of the material structure and device fabrication). High-frequency and synchronous voltage pulses, combined with DC voltages through bias tees, were input to gates $V_1$–$V_6$. Fast dual RF reflectometry was performed by injecting a carrier signal having a frequency $\approx$ 125 MHz (153 MHz) and power of -100 dBm at the Ohmic contacts of the left (right) RF single-electron transistors (see Fig. 1a). The reflected power was monitored through a frequency-multiplexed homodyne detection. The device was operated in a dilution refrigerator with a base temperature $\approx 7$ mK, where an external magnetic field (0.7 T) was applied in the direction shown in Fig. 1a. The measured electron temperature is $\approx 72$ mK(Ref. 28).

The Hamiltonian of the two-qubit system is given (up to a constant term) by[26]

$$H = \frac{J_L(\varepsilon_L)}{2}\sigma_{zL} \otimes I + \frac{\Delta B_{zL}}{2}\sigma_{xL} \otimes I + \frac{J_R(\varepsilon_R)}{2} I \otimes \sigma_{zR} + \frac{\Delta B_{zR}}{2} I \otimes \sigma_{xR} + \frac{J_{RL}(\varepsilon_L,\varepsilon_R)}{2}(\sigma_{zL}-I)\otimes(\sigma_{zR}-I) \quad (1),$$

where $J_L(\varepsilon_L)$ ($J_R(\varepsilon_R)$) is the exchange splitting between states $|S\rangle$ and $|T_0\rangle$, controlled by potential detuning $\varepsilon_L$ ($\varepsilon_R$) of the left (right) qubit ($Q_L$ and $Q_R$). $\sigma_{i=xL,zL}$ ($\sigma_{i=xR,zR}$) is the Pauli matrix for $Q_L$ ($Q_R$), $I$ is the identity matrix, and $\Delta B_{zL}$ ($\Delta B_{zR}$) is the magnetic field difference between the constituent QDs of each qubit, set by the local hyperfine interaction with the host Ga and As nuclei. Here, we adopted $g^*\mu_B/h = 1$ for units, where $g^* \approx -0.44$ is the effective gyromagnetic ratio in GaAs, $\mu_B$ is the Bohr magneton, and $h$ is Planck's constant.

For the real-time Hamiltonian estimation of the quasi-static fluctuations of $\Delta B_{zL}$ and $\Delta B_{zR}$, owing to statistical fluctuations of the nuclei, we extended the methodology developed in Refs. [4,23,28] for two-qubit systems. Briefly, we initialized and measured the $(2,0)_L$ $((0,4)_R)$ charge configuration for $Q_L$ ($Q_R$), respectively, via fast energy selective tunneling (EST)-based single-shot readout and adaptive initialization[29–31] (see also Supplementary Note 1). In contrast to previous studies, in which sequential measurements were performed using the Pauli spin blockade (PSB)-based readout[26,32], here we simultaneously measured both qubits to minimize the estimation latency. During the estimation sequence, when qubits are in the evolution step, $J_L(\varepsilon_L)$, $J_R(\varepsilon_R)$, and $J_{RL}(\varepsilon_L,\varepsilon_R)$ are abruptly turned off. Each qubit evolves around the *x*-axis of the respective Bloch sphere during the evolution time $t_k = 1.67\,k$ ns at the $k^{th}$ estimation trial (Fig. 1b). For each simultaneous single-shot measurement, the Bayesian inference is simultaneously performed according to the following rule (up to a normalization constant)[4]:

$$P(\Delta B_{zi} | m_{Ni}, m_{(N-1)i}, \ldots m_{1i}) = P_0(\Delta B_{zi}) \prod_{k=1}^{N} \frac{1}{2}[1 + r_{ki}(\alpha + \beta \cos(2\pi \Delta B_{zi} t_{ki}))], \quad \text{for } i = \text{L, R}, \quad (2)$$

where $N$ is the number of single-shot measurements per Hamiltonian estimation, $P_0(\Delta B_{zi})$ is the uniform initial distribution, $r_{ki} = 1(-1)$ for $m_{ki} = |S\rangle(|T_0\rangle)$, and $\alpha = 0.1$ ($\beta = 0.8$) is the parameter determined by the axis of rotation on the Bloch sphere (oscillation visibility). After the $N^{th}$ round, the value of $\Delta B_{zi}$ is estimated, where the posterior distribution $P(\Delta B_{zi} | m_{Ni}, m_{(N-1)i}, \ldots m_{1i})$ reaches its maximum. The Bayesian circuit was implemented using a commercial field-programmable gate array (FPGA)[4] (See Supplementary Note 2).

Fig. 1c shows a typical time trace of the estimated $\Delta B_{zL}$ and $\Delta B_{zR}$. The time resolution of these estimations is approximately 1.8 ms, which consists of ($N = 70$) × 26 μs, where a single Bayesian update takes 16 μs and 10 μs for a single-shot measurement and calculation according to Eq. (2), respectively. We found that both $\Delta B_z$s exhibit non-zero average values, likely arising from unintentional nuclear polarization, as reported previously for similarly prepared GaAs devices[17,33]. When ensemble-averaged, this fluctuation leads to a nuclear fluctuation-limited coherence time $T_2^*$ of the order of 20 ns. Moreover, the difference between the mean values of $\Delta B_{zL}$ and $\Delta B_{zR}$ is at least twice the standard deviation. While the microscopic origin of this phenomenon is not well known yet, we used this difference to set the range of qubit frequencies to $25 < \Delta B_{zL} < 50$ ($100 < \Delta B_{zR} < 160$) MHz for simultaneous qubit drive and active frequency feedback.

As shown in the inset of Fig. 2a, we concatenate the Bayesian estimators, acting as probe and operation steps, to control the two qubits in the frequency feedback mode. The controller triggers the operation step when the estimated $\Delta B_z$ in the probe step is in the range described above; otherwise, the cases are discarded. However, the preset range of allowed $\Delta B_z$ covers almost the entire distribution (Fig. 1c), so heralding does not significantly increase the total experimental time. When triggered, each qubit is first adiabatically initialized near the *x*-axis on the Bloch sphere. The controller also adjusts the RF drive frequency using the probed $\Delta B_z$ for each qubit. Then, an RF pulse, with a varied pulse length, is applied to gate $V_2$ ($V_5$) to resonantly modulate $J_L$ ($J_R$) and induce Rabi oscillation. In all experiments, approximately 50 shots of operations were performed after one probe step.

The main panel of Fig. 2a shows the coherent Rabi oscillation for each qubit measured as a function of the RF pulse duration $t_{RF}$ and the controlled detuning $\delta f$ with respect to the actively adjusted resonant frequency. The chevron pattern of $Q_R$ shows a resonance frequency shift, most likely caused by the AC stark effect[34]. Fig. 2b compares the Rabi oscillations for individual and simultaneous qubit operations under resonant conditions. For the former (latter), both the probe and RF pulses were applied to only one qubit (simultaneously on both qubits). For individual operations, $Q_L$ ($Q_R$) shows a Rabi decay time $T_{Rabi} = 1.75$ μs (1.88 μs) at the Rabi frequency $f_{Rabi} = 3.09$ MHz (5.69 MHz) and oscillation visibility of 90.8 % (93.6 %), yielding the oscillation quality factor $Q = f_{Rabi} T_{Rabi} \approx 5$ (11). For simultaneous operations, corresponding results are 1.68 μs (1.59 μs), 3.12 MHz (5.68 MHz), and 88.4 % (88.9 %).

The comparison reveals that the Rabi frequency remains virtually unchanged regardless of the operation scheme, indicating a negligible RF crosstalk between the two qubits. This is expected from the Rabi chevron pattern for each qubit (Fig. 2a) because the separation between the set target frequency range (50 MHz) is larger than the width of each chevron pattern (~10 MHZ). In contrast, simultaneous qubit operation generally reduces $T_{Rabi}$ and oscillation visibility. The time required for the Bayesian calculation depends on the type of operation mode. It takes 10 μs for the single-qubit probe-feedback mode and two-qubit probe-only mode. However, it takes 50 μs for the two-qubit probe-feedback mode, leading to approximately 70 shots × 65 μs = 4.6 ms of latency per Hamiltonian estimation. This is because of the limited resources available on the FPGA setup for parallel computations. Thus, we ascribe the reduced $T_{Rabi}$ to the increased latency for simultaneous Bayesian estimation and frequency feedback, during which the uncertainty of the estimation increases owing to the diffusion of $\Delta B_z$. In addition, reduced oscillation visibility is induced by the finite readout crosstalk between two qubits (see Supplementary Note 3).

Furthermore, we performed a simultaneous Ramsey interference of two qubits by applying calibrated π/2 pulses separated by $t_W$ and by varying $\delta f$, as shown in Fig. 2c. We then extract $T_2^* \approx 151\,(183)$ ns for $Q_L$ ($Q_R$) (inset of Fig. 3c), by fitting the Ramsey amplitude decay to a Gaussian function at each resonant frequency[20]. There is room for further improvements in $T_2^*$ by increasing computational resources of the FPGA setup and removing the readout crosstalk, using newly developed machine learning techniques[35], for example. Nevertheless, we keep these tasks for future work.

**Capacitive coupling between two ST$_0$ qubits**

We now discuss the two-qubit capacitive coupling measurements using the dual Hamiltonian estimation circuit discussed in the previous section. Specifically, throughout the experiment, a resonant RF pulse is applied to the control qubit to observe the state-dependent frequency shift of the target qubit, whose frequency is estimated by the Hamiltonian estimator. In addition, using the simultaneous Hamiltonian estimation circuit, the control qubit is operated when the separation between the qubit frequency is larger than 50 MHz to prevent the unwanted flip of the target qubit by the RF crosstalk.

The capacitive coupling between singlet-triplet qubits originates from the different electric dipole moments of states $|S\rangle$ and $|T_0\rangle$ (Ref. [36]). It has been considered to be a simple method to generate leakage-free two-qubit gates[22,26,27,37] unlike the inter-qubit exchange coupling-based method, in which the inter-qubit magnetic field difference should be sizable to prevent leakage of the qubits outside the computational space[22]. Nevertheless, the weak coupling dependent on the intra-qubit exchange energies constitutes the main disadvantage of the capacitive coupling method. For example, the pioneering demonstration of entanglement in GaAs[26] used a coupling strength on the order of a few MHz, whereas individual exchange energies were approximately 300 MHz. Moreover, it has been assumed that capacitive coupling follows a bilinear form $J_{RL} \propto J_L J_R$. In this bilinear form, the entanglement fidelity is expected to remain constant since the fidelity is limited by the dephasing of an individual qubit in $J_{RL} \ll J_{L(R)}$, giving a constant quality factor for $T_{2L(R)}^* \propto (J_{L(R)})^{-1}$ (Ref. [26]). This constant entanglement fidelity is experimentally confirmed in previous research, indicating that the bilinear form seems to hold, at least for the experiment in which the inter-qubit distance is larger than the distance between dots within a qubit[26,36].

The validity of the previously assumed scaling of $J_{RL}$ was experimentally tested in a regularly and compactly spaced linear QD array. Motivated by theoretical works showing that $J_{RL}$ can actually be a stronger function of $J_L$ and $J_R$ (Ref. [37]), we measured $J_{RL}$ by performing state-dependent exchange oscillation in combination with the dual Hamiltonian estimator. Fig. 3a shows the pulse sequence for the target and control qubits. After the probe step, the control qubit is initialized to the x-axis of a Bloch sphere, followed by an optional π pulse. $J_{RL}$ is then adiabatically switched on by slowly adjusting the detuning of the control qubit while the target qubit is initialized to the x-axis of a Bloch sphere. The exchange oscillation of the target qubit is then performed by diabatically changing the detuning of the target qubit for a time $t_{exch}$ to induce exchange oscillations, whose frequency depends on the control qubit state as $f^T = \sqrt{(J^T - J_{RL} r_C)^2 + \Delta B_z^2}$ according to Eq. (1), where $J^T$ is the intra-qubit exchange energy of the target qubit and $r_C = 0\ (1)$ when the state of the control qubit is $|S\rangle\ (|T_0\rangle)$.

Fig. 3b (Fig. 3c) shows the resultant state-conditional frequency shift of $Q_L$ ($Q_R$) as a function of $t_{exch}$, with $J_{R(L)} \sim 3.61\ (4.13)$ GHz. The precession frequency of the qubit is lower when the control qubit is in the $|S\rangle$ state for both cases, which is consistent with the charge configuration of the QD array[38]. The observed frequency shifts of 34.9 (40.6) MHz for $Q_L$ ($Q_R$) is a direct measure of $J_{RL}$, which is significantly larger than the value reported in Ref.[38]. As predicted in recent theoretical works, we hypothesize that the different relative orientations and the shorter distance between the qubits are related to this enhancement[39]. In addition, we observe the beating of the target qubit oscillation when the control qubit is prepared as a superposition of $|S\rangle$ and $|T_0\rangle$ (see Supplementary Note 4).

We also measured $T_2^*$ and spin-echo coherence time, $T_\text{echo}$, for each qubit to quantify the quality factor of the conditional phase-flip operation. Fig. 4a and 4b show $T_2^*$ and $T_\text{echo}$ for each qubit, where $T_2^*$ is extracted from the exponentially decaying exchange oscillation and $T_\text{echo}$ is measured by fitting the data to the echo envelope using calibrated π/2 and π pulses. Along with the form $J \propto \exp(-\varepsilon)$ (inset of Fig. 4a and 4b), we observe charge noise-limited coherence time, where $T_2^*$ and $T_\text{echo}$ are close to the form $(dJ/d\varepsilon)^{-1}$ (Ref. [20]). This essentially explains why previously demonstrated entanglement quality showed no improvement when increasing $J_L$ and $J_R$ if $J_{RL} \propto J_L J_R$ (Ref. [26]).

Next, we perform the experiment in Fig. 3 with varying $J_L$ and $J_R$ near $J_R = J_L$ to investigate the super-linearity of $J_{RL}$ when both qubits show reasonable coherence. Fig. 4c shows the nonlinear behavior of $J_{RL}$ as $J_L$ and $J_R$ increase, manifesting the deviation from the bilinear proportionality[26,27] in our device. Here, the error bar of the estimated $J_{RL}$ is determined by the fitting uncertainty limited by the sampling rate of the arbitrary waveform generator (see Supplementary Note 5). By fitting the measured $J_{RL}$ to $(J_L J_R)^a$, the agreement with experimental data is found for $a = 2.14$, which is close to the theoretically expected form $J_{RL} \sim (J_L J_R)^2$ using the effective Hamiltonian obtained from a Hund-Mulliken model independent of the details of the confinement potential in the regime where intra-qubit tunnel coupling overwhelms the intra-qubit exchange energy[37] (see Supplementary Note 6). Moreover, we estimated the dipolar energy $D \approx 46$ GHz, the order of which is consistent with the recent experimental work using a similar interdot spacing[40]. With this super-linear proportionality, we observe $J_{RL} > 190$ MHz when $J_L$ and $J_R \approx 900$ MHz, showing that more than 20 % of the state-conditional qubit frequency shift can be obtained in a closely spaced QD array.

**Discussion**

The nonlinear $J_{RL}(J_L, J_R)$ form implies that the two-qubit gate quality should increase at larger $J_L$ and $J_R$. We calculated $Q_{T_2^*(\text{echo})} \equiv 2JT_2^*(T_{\text{echo}})$, which quantifies the number of conditional phase flips within $T_2^*(T_{\text{echo}})$, as shown in Fig. 4d. We observed $Q_{\text{echo}}$ as high as ~16 (~7) for $Q_L$ ($Q_R$), predicting that the fidelity of a conditional phase flip operation of $Q_L$ ($Q_R$) with $Q_R$ ($Q_L$) in the $\sigma_z$ eigenstate reaches as high as $e^{-1/Q_{\text{echo}}} = 94.0$ (86.7) % and also monotonically increases as a function of $J_L$ and $J_R$. In addition, the simulation based on the measured values predicts that the maximum attainable Bell state fidelity $F_{\text{Bell}}$ reaches $\approx 95$ % and increases at larger $J_L$ and $J_R$, where the Bell state is prepared by the echo-like pulse implemented in Ref.[26] in which $F_{\text{Bell}}$ maximizes at $\approx 72\%$ with $J_{RL} \approx 1\,\text{MHz}$ (See Supplementary Note 7).

Previously, a Bell state fidelity with a capacitive coupling has enhanced to $\approx 93\%$ with simultaneous rotary echo and rapid dynamic nuclear polarization (DNP), which enabled an approximately 10-fold increase of a coherence time with $\Delta B_z = 900$ MHz, but $J_{RL}$ on the order of a few MHz was rather exploited[27]. Thus we expect that $F_{\text{Bell}}$ could be enhanced more by applying simultaneous rotary echo to the closely spaced QD array at a large $J_{RL}$, although the application is not currently viable in our device due to the insufficient DNP rate. In addition, the minimum time step and pulse rise times are currently limited by the sample rate of the waveform generator (2.4 Gsa/s), which also prevents performing full two-qubit gate operations and entanglement demonstration. Therefore further optimization with a faster signal source is still required. Note also that the pulse sequence in Fig. 3a is proper only for a two-qubit interaction measurement since the control qubit is likely to decohere while adiabatically turning on the interaction. Thus, a different qubit driving strategy (for example, using a non-adiabatic pulse) should be devised for entanglement demonstration, which will be considered in future work. Nonetheless, as our Hamiltonian estimation technique and readout method are compatible with large $\Delta B_z$, we anticipate that performing a full two-qubit experiment in a regularly and closely spaced linear QD array, with increased $\Delta B_z$ by micromagnets or dynamic nuclear polarization, may show an even higher two-qubit gate fidelity that is also fast, exploiting large $J_{RL}$.

In conclusion, we demonstrated the simultaneous Hamiltonian parameter estimation and active suppression of the quasi-static noise of two $ST_0$ qubits in a GaAs quadruple QD array. Using fast qubit calibration routines, we also showed that both the magnitude and scaling of the capacitive coupling in a closely spaced QD array can be stronger than the previously measured bilinear form, leading to a state-conditional frequency shift of over 20 % and a quality factor of conditional phase flip of over 16. Our measurement confirms recent theoretical calculations and supports the possibility of realizing a high-fidelity and fast entanglement of encoded spin qubits in both GaAs and Si using a simple capacitive interaction.

**Methods**

**Device Fabrication.** The quadruple QD device shown in Fig. 1a was fabricated on a GaAs/AlGaAs heterostructure where two-dimensional electronic gas (2DEG) is located 70 nm below the surface. Mesa was defined by a wet etching technique to eliminate 2DEG outside the region of interest to suppress unwanted leakage. Five ohmic contacts were formed by metal diffusion with thermal annealing. Nano-gates were fabricated by e-beam lithography and metal evaporation.

**Measurement.** The device was placed on the 7 mK plate in a commercial dilution refrigerator (Oxford Instruments, Triton-500). The battery-operated voltage sources (Stanford Research Systems, SIM928) supplied by stable DC voltages rapid voltage pulses generated by the arbitrary waveform generator with the maximum sampling rate of 2.4Gsa/s (Zurich Instruments, HDAWG) were applied to metallic gates through on-board bias-tees. A detailed description of the experimental setup and FPGA implementation can be found in Supplementary Note 2.

**Data Availability**

The data that support the findings of this study are available from the corresponding author upon request.

**Acknowledgments**

This work was supported by the National Research Foundation of Korea (NRF) grant funded by the Korean Government (MSIT) (No. 2018R1A2A3075438, No. 2019M3E4A1080144, No. 2019M3E4A1080145, and No. 2019R1A5A1027055), Korea Basic Science Institute (National Research Facilities and Equipment Center) grant funded by the Ministry of Education (No. 2021R1A6C101B418), and Creative-Pioneering Researchers Program through Seoul National University (SNU). The cryogenic measurement used equipment supported by the Samsung Science and Technology Foundation under Project Number SSTF-BA1502-03. Correspondence and requests for materials should be addressed to DK (dohunkim@snu.ac.kr).

**Figure Legends**

**Figure 1. Experimental setup. a.** A schematic of the experimental setup. Yellow circles indicate the RF single-electron transistors for dual RF-reflectometry. Orange (green) circles indicate QDs for the left (right) $ST_0$ qubit ($Q_L$ and $Q_R$). The plunger and barrier gates are connected to the arbitrary waveform generator for the application of detuning and RF pulses. The white scale bar corresponds to 500 nm. **b.** Schematic diagram of the dual Hamiltonian estimation of the field gradient $\Delta B_L$ and $\Delta B_R$. See main text for the estimation procedure. The green box shows the energy selective tunneling-based single-shot readout method. **c.** The example traces of the simultaneously estimated field gradient $\Delta B$ of each qubit as a function of time.

**Figure 2. Simultaneous driving of two $ST_0$ qubits. a.** Individually driven Rabi oscillation of the triplet return probability $P_T$ of $Q_L$ and $Q_R$ as a function of controlled detuning $\delta f$ and RF pulse duration $t_{RF}$. **b.** The representative $P_T$ oscillation of each qubit as a function of RF pulse duration for an individual and simultaneous operation. Solid curves are a fit to sinusoidal functions with a Gaussian envelope. **c.** Ramsey fringe of $P_T$ as a function of controlled detuning $\delta f$ and Ramsey delay $t_W$, showing typical Ramsey interference pattern. Inset: The line cut at the resonance condition. The solid curve is a fit to a Gaussian decay function $P_T(t_W) = A\exp\left(-(t_W/T_2^*)^2\right) + B$ with the best fit parameter $T_2^* \approx 151$ (183) ns for $Q_L$ ($Q_R$).

**Figure 3. Demonstration of inter-qubit coupling. a.** A pulse sequence to produce the control qubit ($Q_C$) state-dependent exchange oscillation of the target qubit ($Q_T$). A schematic illustration of the state of both qubits in each step is depicted in the upper panel. **b.(c.)** The exchange oscillations of $Q_L$ ($Q_R$) depending on the state of the other qubit as control. The lower and upper traces are measured when the state of $Q_T$ is $|S\rangle$ and $|T_0\rangle$, respectively.

**Figure 4. Super-linearly increasing inter-qubit coupling strength. a.(b.)** $T_2^*$ and $T_{echo}$ of $Q_L(Q_R)$. The solid curves are fit to the forms $(dJ/d\varepsilon)^{-b}$ with best fit parameter $b \approx 1.0$. Inset: the intra-qubit exchange energy of each qubit as a function of detuning amplitude $\varepsilon$. The solid line denotes an exponential fit of a form $J(\varepsilon) = J_0 + J_1 \exp((\varepsilon_0 - \varepsilon)/\alpha)$. **c.** $J_{RL}$ (black circle) and $J_{RL}/J_L J_R$ (red square) as a function of $J_L$ and $J_R$. The black solid line is a fit to $J_{RL} = a(J_L J_R)^\alpha$ with the best fit parameter $\alpha = 2.14$ indicating super-linear behavior. The error bars are estimated from the fitting uncertainty. **d.** The conditional phase flip quality factors $Q_{T_2^*}$ and $Q_{echo}$ of $Q_L$ (orange circle) and $Q_R$ (green circle).

Figure 1

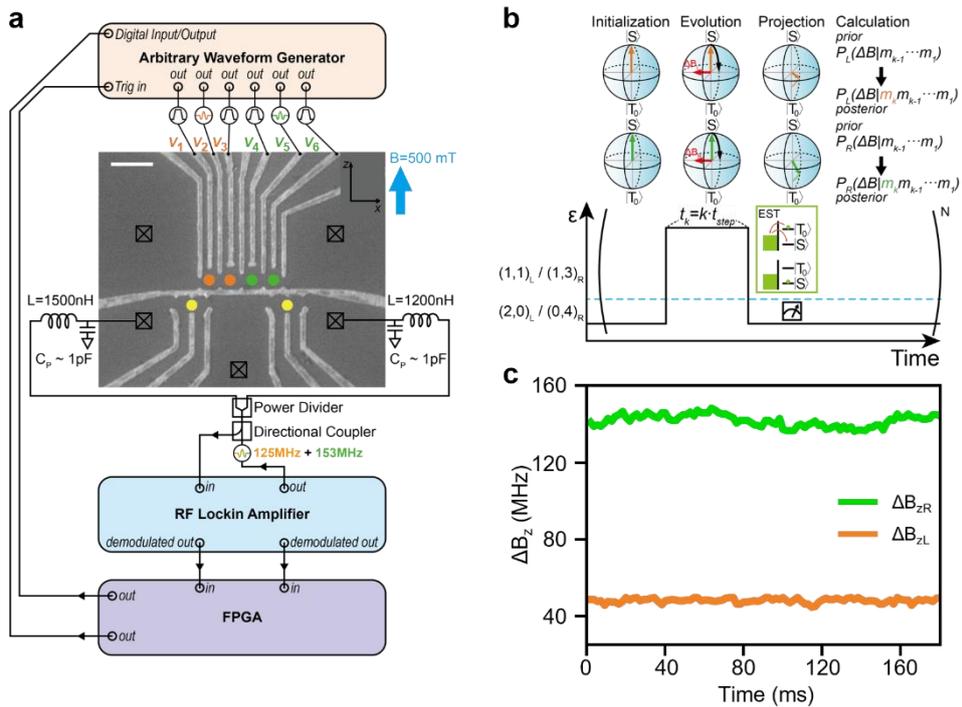

Figure 2

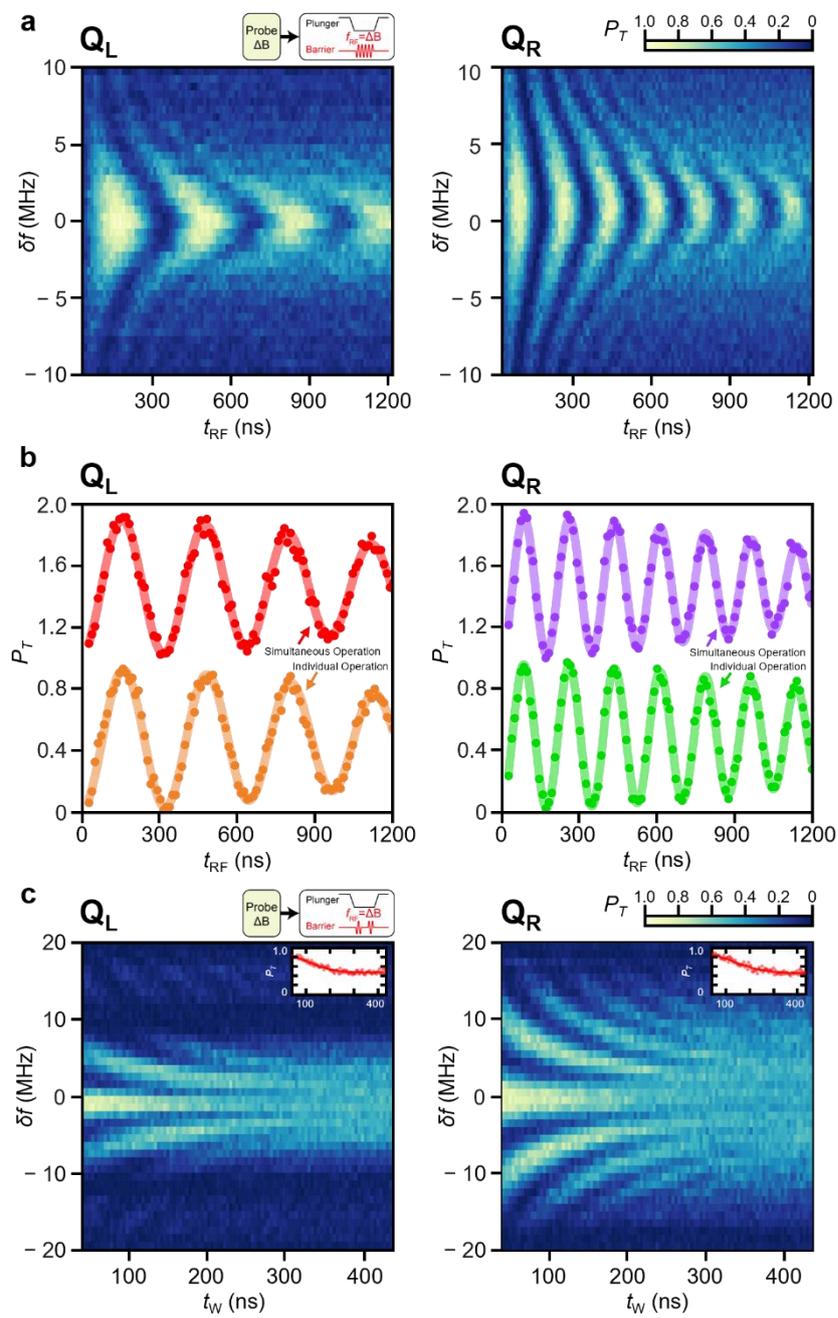

Figure 3

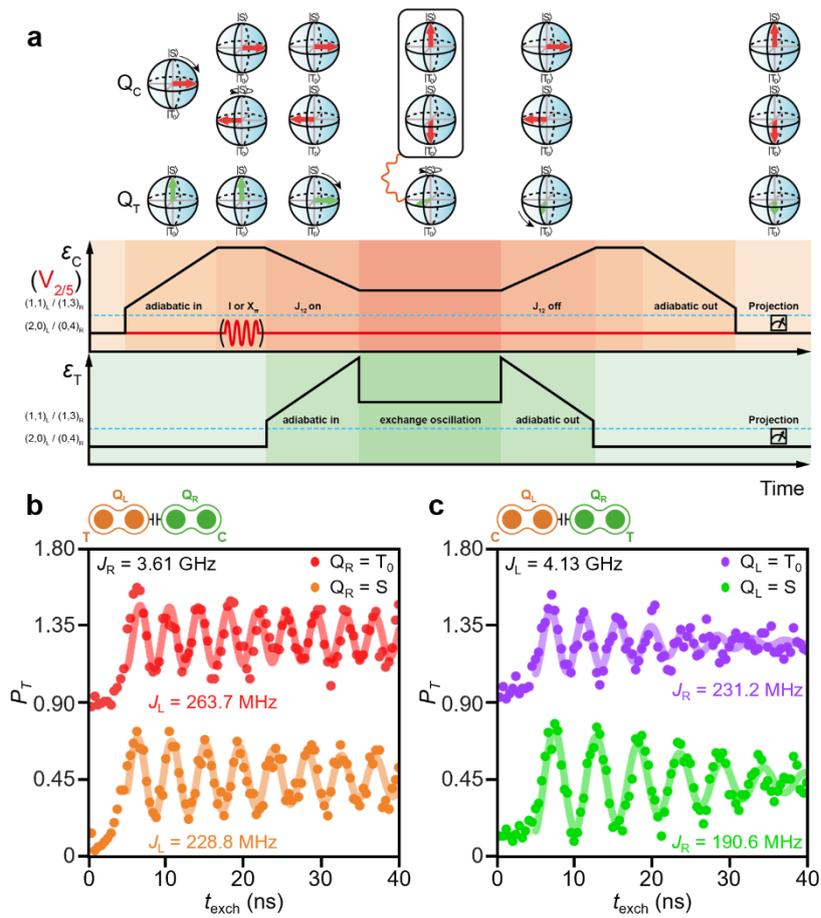

Figure 4

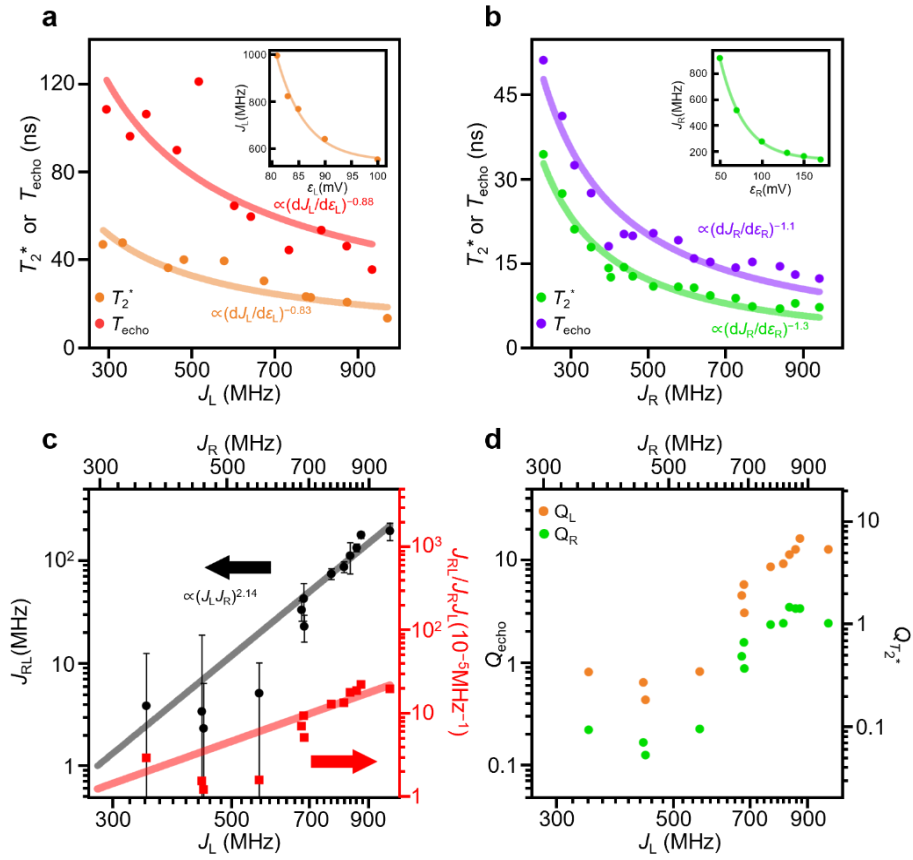

**Supplementary Materials**

**Supplementary Note 1. Charge Stability Diagrams**

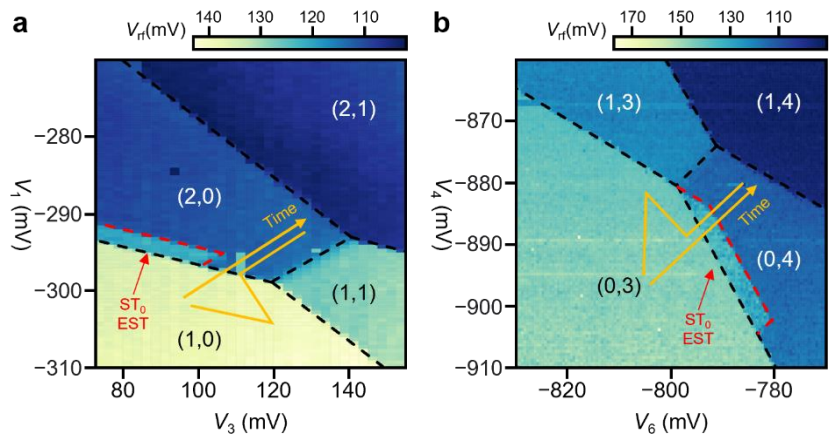

**Supplementary Figure 1. Charge stability diagram of the left and right double quantum dots. a. (b.)** Charge stability diagram of the left (right) double quantum dot measured with the triangular pulse superimposed with raster scanning gate voltages. The red dashed lines depict the boundaries of the energy selective tunneling (EST) readout window.

Supplementary Fig. 1 shows the charge stability diagrams of the double quantum dots used in the two-qubit experiments. The diagrams were measured by superposing the raster scanning voltage with the triangular pulse depicted in each figure, with a rise-in (-out) time of ~400 ns (~400 ps). Near the readout window (red dashed lines), the triangular pulse adiabatically brings the initial-state $|S\rangle$ to the polarized triplet $|T_+\rangle$ using ST$_+$ anticrossing and returns to the initial detuning abruptly. The resulting $|T_+\rangle$ state rapidly tunnels out to the reservoir, inducing transient RF sensor responses. The time-averaged measurement of tunneling events inside the readout window was used to calibrate the optimal qubit readout positions. Note that the (0,4)–(1,3) charge configuration is used to define the right ST$_0$ qubit. Exploiting the screening effect of the core-shell electron pairs, we used the increased singlet-triplet energy splitting and the readout window of the right qubit[1–3].

**Supplementary Note 2. Experimental Setup and FPGA Implementation**

The bulk of the experimental setup used in this study is described in Supplementary Ref.[4]. It has been extended for the simultaneous measurement and operation of two qubits. Two RF-single-electron transistor (RF-SET) sensors, with impedance matching tank circuits for a resonance frequency of ~ 125 MHz (~ 153 MHz), were implemented for the charge sensing of the left (right) singlet-triplet ($ST_0$) qubit. A commercial high-frequency lock-in amplifier (Zurich Instrument, UHFLI) was used as the carrier generator, which sends a dual-tone carrier with a power of −40 dBm, which is further attenuated by −60 dB inside the cryostat, and frequencies of 125 MHz and 153 MHz to two Ohmic contacts via a directional coupler and power divider. The reflected signals from each RF-SET were first combined and amplified by 50 dB using cryogenic amplifiers (Cosmic Microwave Technology, CITLF2 × 2 in series) at 4 K. The resultant signal was further amplified at room temperature by 25 dB using a custom-built low-noise RF amplifier.

For real-time data processing, we implemented a digital logic circuit with a field-programmable gate array board (FPGA; Digilent Zedboard with Zynq-7000 XC7Z020-CLG484). Two RF-demodulated analog signals from the UHFLI were input to the two 12-bit analog-to-digital converter channels of the FPGA. For single-shot discrimination, the transient tunneling events of each qubit state were thresholded in real-time by comparing the preset threshold values with the data in parallel. The discriminator records bit 1 immediately when the data above the threshold value are detected. Bit 0 was recorded when such events did not occur throughout the preset measurement period of the 15 $\mu s$.

After the single-shot discrimination of both qubits, the FPGA follows either of the following steps depending on the operation mode. First, for the individual active feedback operation mode (Fig. 3a of the main text), the Bayesian estimation of the qubit frequency is performed by computing the posterior probability distribution between 0 and 100 MHz (70 and 170 MHz) for the left (right) qubit. We use a look-up table (LUT) that stores all possible values of the likelihood function in the block RAM inside the FPGA and design a 512-parallelized calculation module to minimize latency due to data processing. The FPGA then converts the estimated frequency to a 9-bit digital signal and sends it to the digital input/output port of the arbitrary waveform generator (Zurich Instruments, HDAWG), provided that the estimated qubit frequency is within the preset range. The HDAWG then applies a square-wave enveloped waveform with a frequency corresponding to the digital value $V_2$ ($V_5$) for the left (right) qubit operation using the multifrequency modulation function.

Second, for simultaneous Hamiltonian estimation without the operation step (Fig. 1c of the main text), two Bayesian update circuits inside the FPGA were run simultaneously. Up to this operation mode, the FPGA supports sufficient computational resources for implementing 512 parallelized calculation modules, and the latency for one Bayesian update after each single-shot measurement is approximately 10 μs.

Third, for simultaneous Hamiltonian estimation and simultaneous frequency feedback operation mode (Fig. 2b, 2c, 3, and 4 of the main text), the FPGA runs two copies of modules described for the first operation mode. Owing to the lack of computational resources, the number of parallelized modules was reduced to 128, and the resulting latency was approximately 50 μs for this operation mode. This is the main reason for the reduced Rabi decay time, $T_{\text{Rabi}}$ observed in Fig. 2b of the main text. In all steps, an adaptive state initialization is conducted by acquiring a 200 ns long sample and thresholding repeatedly until the last value falls below the threshold after 15 μs of the preset single-shot measurement time.

**Supplementary Note 3. Readout Crosstalk**

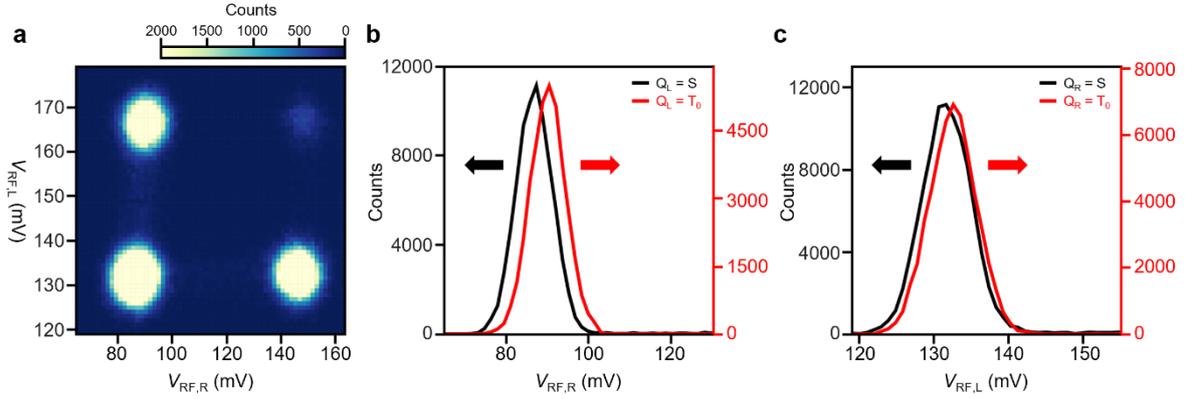

**Supplementary Figure 2. Readout crosstalk analysis. a.** A two-dimensional histogram of the left ($V_{RF,L}$) and the right ($V_{RF,R}$) RF-SET response. **b.(c.)** Histogram of the right (left) RF-SET response. The black (red) curve is obtained when the other qubit is $|S\rangle$ ($|T_0\rangle$), showing non-negligible readout crosstalk.

As RF-SETs are capacitively coupled to all four dots comprising the left ($Q_L$) and right ($Q_R$) $ST_0$ qubits, the sensor response also depends on the charge configuration of the remote $ST_0$ qubit. In conjunction with the transient (0,1) or (0,3) charge configuration introduced by tunneling out of the $|T_0\rangle$ state, this dependency should result in the state-dependent crosstalk effect of the RF-SET response in the EST readout point[5].

Supplementary Fig. 2a shows a two-dimensional histogram of the responses of the left ($V_{RF,L}$) and right ($V_{RF,R}$) RF sensors at the EST readout point. Here, the response of the left (right) sensor was obtained by demodulating the reflected RF signal at 125 MHz (153 MHz). The location of the histogram peak forms a parallelogram, which shows slight obliqueness caused by the state-dependent crosstalk effect. In addition, the peak shift of the $V_{RF,R}$ ($V_{RF,L}$) histogram, illustrated in Supplementary Fig. 2b (c) reveals the state-dependent crosstalk effect of the right (left) RF-SET sensor. Notably, the peak shift of $V_{RF,R}$ histogram is 3.14 mV, which is greater than that of $V_{RF,L}$ histogram by 2.18 mV. Because the visibility decrease under simultaneous operation is also greater for $Q_R$ ($Q_R$:4.7 %, $Q_L$:2.4 %), we ascribe the decrease in the oscillation visibility under the simultaneous operation to the readout crosstalk effect.

**Supplementary Note 4. Additional Interqubit Coupling Analysis**

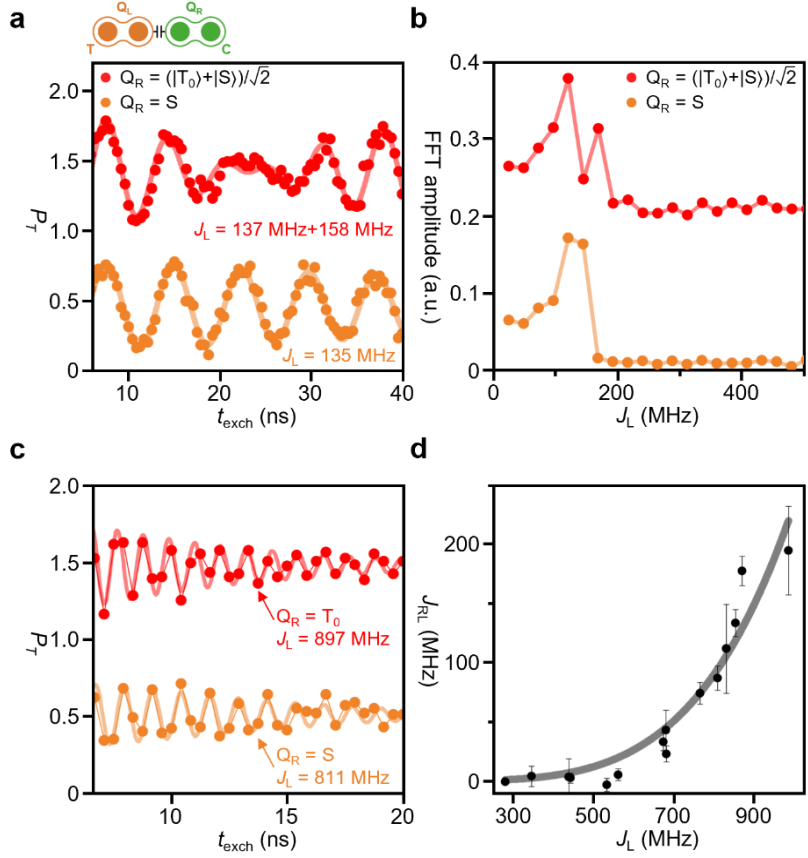

**Supplementary Figure 3. Interqubit coupling analysis. a.** The example exchange oscillation of $Q_L$ as a function of exchange duration $t_{exch}$ depending on the state of $Q_R$. The solid curve denotes a curve fit to the measured data, which yields $J_L = 135$ MHz (above) and $J_L = 137$ MHz+158 MHz (below). **b.** FFT spectra of $J_L$ obtained from (a). The peak splitting is observed when the state of $Q_R$ is the superposition of S and $T_0$. **c.** The exchange oscillation of the left qubit ($Q_L$) with varied right qubit ($Q_R$) state as a function of exchange duration $t_{exch}$. The transparent solid line is a curve fit to the corresponding data. **d.** $J_{RL}$ and its uncertainty as a function of $J_L$. The data is plotted on a linear scale to clarify the tendency of the uncertainty. The solid black line is a fit of $J_{RL} = a(J_L J_R)^\alpha$ to the measured data.

Supplementary Fig. 3a shows the exchange oscillations of target qubit $Q_L$ depending on the state of control qubit $Q_R$, measured using the pulse sequence in Fig. 3a of the main text.

Here, instead of preparing $Q_R$ into $T_0$, we prepare the superposition of S and $T_0$ state by applying a π/2 gate instead of a π gate. As expected for a coupled two qubit system, the oscillation shows beating behavior when $Q_R$ is in the superposed state. By fitting the data to the formula $A(\cos(2\pi f_1 t_{exch} + \varphi) + \cos(2\pi f_2 t_{exch} + \varphi))\exp(-(t_{exch}/T_2^*)^a)) + B$, we measure $J_L$ to be superposition of oscillations with frequencies of 137 MHz and 158 MHz. Furthermore, the fast-Fourier transformation (FFT) of the oscillation illustrated in Supplementary Fig. 3b shows the FFT amplitude peak at 120 MHz splitting into the peaks at 130MHz and 160 MHz when $Q_R$ changed from S to the superposition of S and $T_0$ state, suggesting the interference between two different qubit frequencies. Along with the frequency shift depending on the control qubit state illustrated in Fig. 3c in the main text, the beating of the oscillation confirms the coupling between two $ST_0$ qubits.

Supplementary Fig. 3c shows an example of exchange oscillations of $Q_L$ with different $Q_R$ states measured by the pulse sequence in Fig. 3a of the main text. The intra-qubit exchange energy of each qubit $J_{i=R,L}$ was extracted by fitting the formula $A\cos(2\pi f t_{exch} + \varphi)\exp(-(t_{exch}/T_2^*)^a)) + B$ to the measured exchange oscillation. The $J_{RL}$ at each qubit frequency was then obtained from the oscillation frequency shift of the target qubit under different control qubit states. From the fitting uncertainty $\sigma_J$ ($\sigma_J'$) of the target qubit frequency when the control qubit is S ($T_0$), we calculated the uncertainty of $J_{RL}$ as $\sigma_{RL} = \sqrt{\sigma_J^2 + \sigma_J'^2}$ by considering the propagation of error. Similar to the Fig. 4c of the main text, we plot the estimated $J_{RL}$ with $\sigma_{RL}$ in Supplementary Fig. 3d, but here we plot in a linear scale to clarify the tendency of the uncertainty of $J_{RL}$.

**Supplementary Note 5. Extraction of Exchange Energies by Curve Fitting**

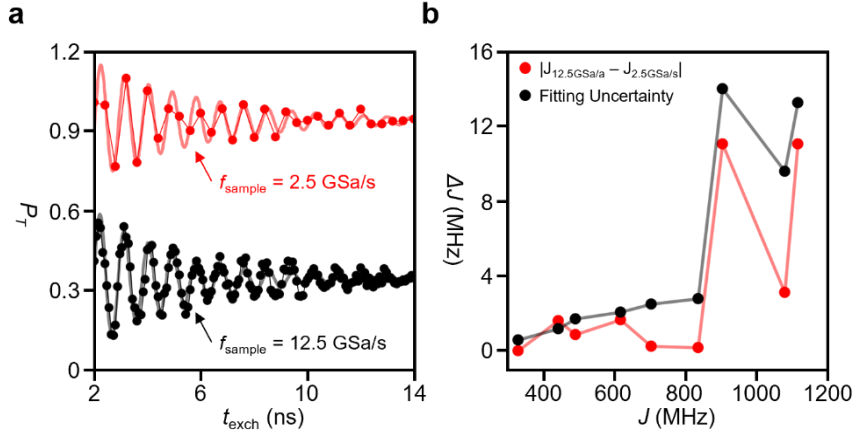

**Supplementary Figure 4. Curve fitting uncertainty and the limited sampling rate of arbitrary waveform generator. a** The exchange oscillation of $ST_0$ qubit as a function of exchange duration $t_{exch}$ sampled at two different sampling rates $f_{sample}$ (red: 2.5 GHz, black: 12.5 GHz). The solid curve denotes a curve fit to the measured data, which yields $f = 1116.15 \pm 2.7$ ($1105.15 \pm 13.26$) MHz for $f_{sample} = 12.5$ (2.5) GSa/s . **b.** The fitting uncertainty at the sampling rate of 2.5 GHz (black) and the difference between the estimated qubit frequency $|J_{12.5\,GSa/s} - J_{2.5\,GSa/s}|$ at the sampling rate of 2.5 GSa/s and 12.5 GSa/s (red). Both quantities tend to increase with the increasing qubit frequency $J$.

As described in the main text, we used the waveform generator with maximum sampling rate of 2.4 GSa/s. To show the reliability of the curve fitting procedure, we compare fitting uncertainty with control experiments using a faster waveform generator. Supplementary Fig. 4a shows the exchange oscillation data at the different sampling rates (top: 12.5 GSa/s, bottom: 2.5 GSa/s), which are operated in the same detuning. Here, the data is obtained by a high sampling rate arbitrary waveform generator (Keysight Technologies, M8195A). The curve fitting yields $1116.15 \pm 2.7$ MHz ( $1105.15 \pm 13.26$ MHz ) for the sampling rate of 12.5 (2.5) GSa/s. Although there is a slight difference between the estimated qubit frequency at the sampling rate of 12.5 GSa/s and 2.5 GSa/s, the difference is comparable to the fitting uncertainty at 2.5 GSa/s.

Supplementary Fig. 4b plots the fitting uncertainty of qubit frequency at the sampling rate of 2.5 GSa/s and the difference of estimated qubit frequency $|J_{12.5\text{ GSa/s}} - J_{2.5\text{ GSa/s}}|$ at various qubit frequencies. The tendency in Supplementary Fig. 4b shows that the frequency difference $|J_{12.5\text{ GSa/s}} - J_{2.5\text{ GSa/s}}|$ is comparable to the fitting uncertainty, revealing the validity of curve fitting within the fitting uncertainty below the Nyquist frequency of 2.5 GSa/s. These results shows that the uncertainty of the exchange energies measured in the current work is not limited by the sampling rate of the waveform generator. Since M8195A, in spite of its high sampling rate, has significant limitation for adaptive multi-channel operation and maximum output amplitude, and we choose to use HDAWG in this work.

**Supplementary Note 6. Asymptotic Formula for the Interqubit Capacitive Coupling**

We derive the form of $J_{RL} \propto (J_L J_R)^2$ using the Hamiltonian considered in Supplementary Ref. [6] for two singlet-triplet qubits, which equals the Hamiltonian of Eq. (1) of the main text up to a constant matrix.

$$H = \begin{pmatrix} 0 & 0 & 0 & 0 \\ 0 & -J_R(\varepsilon_R) & 0 & 0 \\ 0 & 0 & -J_L(\varepsilon_L) & 0 \\ 0 & 0 & 0 & E_{SS} \end{pmatrix}. \quad (1)$$

The interqubit capacitive coupling $J_{RL}$ is defined by the quantity $E_{SS}$ using $J_{RL} = E_{SS} + J_L + J_R$, which is the lowest energy eigenvalue of the following Hamiltonian:

$$H_{SS} = \begin{pmatrix} 0 & t_R & t_L & 0 \\ t_R & -J_R + \frac{t_R^2}{J_R} & 0 & t_L \\ t_L & 0 & -J_L + \frac{t_L^2}{J_L} & t_R \\ 0 & t_L & t_R & -J_L - J_R + \frac{t_R^2}{J_R} + \frac{t_L^2}{J_L} + D \end{pmatrix}, \quad (2)$$

where $t_{i=L,R}$ ($J_{i=L,R}$) denotes the intra-qubit tunneling (exchange) energy of each qubit, and $D$ is the dipolar energy between the two qubits. We approximate $E_{SS}$ perturbatively under the following conditions: First, the diagonal terms of the Hamiltonian $H_{SS}$ overwhelm the off-diagonal terms in the regime of interest, that is $J_L \sim J_R \ll t_L \sim t_R \sim D$. Second, the lowest eigenvalue $E_{SS}$ in the $J_L \sim J_R \ll t_L \sim t_R \sim D$ regime should return to zero during $J_L = J_R = 0$. The resultant perturbative expansion of $E_{SS}$ can be written as:

$$E_{SS} = -\frac{2 J_L^2 J_R^2 + J_L^3 J_R \left(\frac{t_R^2}{t_L^2}\right) + J_R^3 J_L \left(\frac{t_L^2}{t_R^2}\right)}{t_L^2 J_R + t_R^2 J_L} + D \frac{2 J_L^3 J_R^3 + J_L^4 J_R^2 \left(\frac{t_R^2}{t_L^2}\right) + J_R^4 J_L^2 \left(\frac{t_L^2}{t_R^2}\right)}{\left(t_L^2 J_R + t_R^2 J_L\right)^2} + \frac{J_R^2 J_L}{t_R^2} + \frac{J_L^2 J_R}{t_L^2} - J_R - J_L + O\left(\left(\frac{J}{t}\right)^5\right). \quad (3)$$

The simple algebraic manipulation of Eq. (3) with the condition $J_L \sim J_R$ yields $J_{RL}$ of the form:

$$J_{RL} = D \frac{(J_L J_R)^2}{(t_L t_R)^2} + O\left(\left(\frac{J}{t}\right)^5\right). \quad (4)$$

Using the measured values of $t_\mathrm{L} = 11.9$ GHz and $t_\mathrm{R} = 3.2$ GHz for the current device, the fit given in Fig. 4c of the main text yields $D = 46$ GHz, the order of magnitude of which is consistent with the value reported in a recent study[7]. Note that for $J_\mathrm{L} \sim J_\mathrm{R} \ll t_\mathrm{L} \sim t_\mathrm{R} \sim D$, the sign of the dipolar energy $D$ does not affect $J_\mathrm{RL}$ significantly.

**Supplementary Note 7. Maximum Attainable Bell State Fidelity**

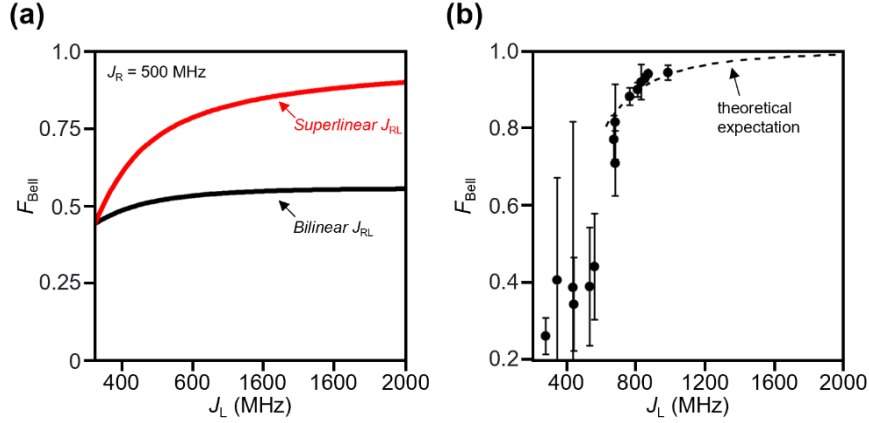

**Supplementary Figure 5. Maximum attainable Bell state fidelity. a** The simulated maximum attainable Bell state fidelity as a function of $J_L$ for bilinear (black) and superlinear (red) $J_{RL}$. The fidelity is estimated at $J_R = 500$ MHz. **b.** The estimated maximum attainable Bell state fidelity from the experimental values. The fidelity shows an increasing tendency with a maximum value of $\approx 0.95$. The error bars are estimated from the uncertainty of $J_{RL}$ and $T_{echo}$ of each qubit.

In this section, we calculate the maximum attainable Bell state fidelity $F_{Bell}$ by simulating the operation using the echo-like pulse implemented in Supplementary Ref.[5] under single qubit dephasing. The generalized Bell state was prepared in the simulation by applying the operation sequence $(X_{\pi/2} \otimes X_{\pi/2})ZZ'(\pi/2J_{RL})(X_\pi \otimes X_\pi)ZZ'(\pi/2J_{RL})$ to the initial state $|S\rangle \otimes |S\rangle$, where $ZZ'(t) = diag(e^{-i(J_L+J_R)t/2}, e^{-i(J_L-J_R)t/2}, e^{i(J_L-J_R)t/2}, e^{i(J_L+J_R)t/2 - iJ_{RL}t})$ is the simultaneous $\sigma_z$ rotation in the presence of inter-qubit capacitive coupling $J_{RL}$. In the absence of a dephasing effect, this operation sequence with $t = \pi/2J_{RL}$ yields the generalized Bell state $e^{i\pi(\sigma_y \otimes I + I \otimes \sigma_y)}(|SS\rangle - |T_0 T_0\rangle)/\sqrt{2}$. The maximum attainable Bell state fidelity $F_{Bell}$ is estimated from the fidelity between the dephasing-free ideal generalized Bell state and the dephasing-included final state.

Supplementary Fig. 5a shows $F_{Bell}$ with $J_{RL}$ of bilinear and superlinear dependence as a function of $J_L$ with $J_R$ fixed at 500 MHz. For the $F_{Bell}$ estimation, the superlinear $J_{RL}$ is calculated using Eq. (3), and dephasing time is obtained from the fitted curve of Fig. 4a and 4b of the main text. Notably, $F_{\text{Bell}}$ of the superlinear $J_{\text{RL}}$ shows an increasing tendency, with the rate being faster than that of the bilinear $J_{\text{RL}}$.

We next estimate $F_{\text{Bell}}$ for $J_{\text{L}} \sim J_{\text{R}}$ from the experimental values. Supplementary Fig. 5b shows $F_{\text{Bell}}$ estimated from the experimentally measured values in Fig. 4 of the main text. Specifically, $F_{\text{Bell}}$ is estimated with the dephasing of both qubits, whose characteristic time was determined from the curve fitted to the $T_{\text{echo}}$ data of Fig. 4a and 4b in the main text. As expected, $F_{\text{Bell}}$ increases with $J_{\text{L}}$ and $J_{\text{R}}$, reaching 94.5±1.9% in the experimentally probed regime. In addition, the simulation with a single-qubit dephasing effect expects that $F_{\text{Bell}}$ increases with $J_{\text{L}}$ and $J_{\text{R}}$, as indicated by the dashed line in Supplementary Fig. 5b. Although the two-qubit dephasing effect becomes important as $J_{\text{RL}}$ increases and could reduce $F_{\text{Bell}}$ from our current estimate, we expect that the entanglement fidelity between two $ST_0$ qubit could be enhanced at the superlinear $J_{\text{RL}}$ regime. More accurate characterization of the entanglement fidelity requires detailed features of the two-qubit dephasing effect, including its characteristic time and noise characteristics.